\documentclass[journal]{IEEEtran}

\ifCLASSINFOpdf
\usepackage[pdftex]{graphicx}
\else
\usepackage[dvips]{graphicx}
\fi
\usepackage{bbm}
\usepackage{url}
\usepackage{tabularx}
\usepackage{subfig}
\usepackage{colortbl,booktabs}
\usepackage{multirow}
\usepackage{threeparttable}
\usepackage{makecell}
\usepackage{algorithm}
\usepackage{verbatim}
\usepackage{algorithmic}
\usepackage{xcolor}
\usepackage{diagbox}
\usepackage{nomencl}
\usepackage{hyperref}
\usepackage{caption}
\makenomenclature

\usepackage{amsthm,amsmath,amssymb}

\usepackage{etoolbox}
\renewcommand\nomgroup[1]{%
  \item[\bfseries
  \ifstrequal{#1}{R}{Random variables}{%
  \ifstrequal{#1}{F}{Functions}{%
  \ifstrequal{#1}{V}{Variables}{%
  \ifstrequal{#1}{M}{Models}{}}}}%
]}

\begin{document}

\title{Continuous and Distribution-free Probabilistic Wind Power Forecasting: A Conditional Normalizing Flow Approach}

\author{
Honglin Wen,~\IEEEmembership{Student~Member,~IEEE,}
Pierre Pinson,~\IEEEmembership{Fellow,~IEEE,}
        Jinghuan~Ma,
        Jie Gu,
        and~Zhijian~Jin% <-this % stops a space
        
\vspace{-2em}
\thanks{
Honglin Wen, Jinghuan Ma, Jie Gu, and Zhijian Jin are with Department of Electrical Engineering, Shanghai Jiao Tong University.\\
Honglin Wen, Pierre Pinson are with the Technical University of Denmark, Department of Technology, Management and Economics.}}

\markboth{IEEE Transactions on Sustainable Energy}%
{Shell \MakeLowercase{\textit{et al.}}: Bare Demo of IEEEtran.cls for IEEE Journals}

\maketitle

\begin{abstract}
We present a data-driven approach for probabilistic wind power forecasting based on conditional normalizing flow~(CNF). In contrast with the existing, this approach is distribution-free (as for non-parametric and quantile-based approaches) and can directly yield continuous probability densities, hence avoiding quantile crossing. It relies on a base distribution and a set of bijective mappings. Both the shape parameters of the base distribution and the bijective mappings are approximated with neural networks. Spline-based conditional normalizing flow is considered owing to its non-affine characteristics. Over the training phase, the model sequentially maps input examples onto samples of base distribution, given the conditional contexts, where parameters are estimated through maximum likelihood. To issue probabilistic forecasts, one eventually maps samples of the base distribution into samples of a desired distribution. Case studies based on open datasets validate the effectiveness of the proposed model, and allows us to discuss its advantages and caveats with respect to the state of the art.
\end{abstract}

% Note that keywords are not normally used for peerreview papers.
%\vspace{-0.5em}

\IEEEpeerreviewmaketitle

\mbox{}

\nomenclature[R]{\(Y_{i,t}\)}{The random variable for wind power generation value at wind farm $i$ at time $t$}
\nomenclature[R]{\(\boldsymbol Y_t\)}{The random variable for wind power generation values in general form at time $t$}
\nomenclature[R]{\(\boldsymbol Z_t\)}{The intermediate random variable at time $t$}
\nomenclature[F]{\(f_{Y_{i,t}}(\cdot)\)}{Probability density function of $Y_{i,t}$}
\nomenclature[F]{\(q^{(\alpha)}(\cdot)\)}{Quantile function with level $\alpha$}
\nomenclature[F]{\(\phi(\cdot)\)}{The function that estimates the shape parameters of base distribution}
\nomenclature[F]{\(c_k(\cdot)\)}{The conditioner function in the $k$-th transform that outputs the conditionals}
\nomenclature[F]{\(\tau_k(\cdot)\)}{The transformer function in the $k$-th transform that maps $z_{t,i}^{(k-1)}$ to $z_{t,i}^{(k)}$}
\nomenclature[F]{\(T_k(\cdot)\)}{The function that maps $\boldsymbol z_t^{(k-1)}$ to $\boldsymbol z_t^{(k)}$}
\nomenclature[M]{\(\mathcal{M}\)}{The whole model}
\nomenclature[M]{\(\mathcal{G}\)}{The model for base distribution}
\nomenclature[V]{\(\boldsymbol x_t\)}{The input features at time $t$}
\nomenclature[V]{\(\boldsymbol y_t\)}{The realization of $\boldsymbol Y_t$}
\nomenclature[V]{\(\boldsymbol z_t\)}{The realization of $\boldsymbol Z_t$}

\printnomenclature

%\vspace{-1em}
\section{Introduction}

\subsection{Motivation}
As an essential tool to assess and accommodate wind power generation uncertainty, short-term probabilistic wind power forecasting (PWPF) has gained increasing interest in recent decades. It generally takes numerical weather prediction and historical values as input features, in order to model and communicate the probability density of wind power generation at some time in the future. Such densities may be for a unique lead time and location (hence, univariate), or jointly for several lead times and/or locations (referred to as multivariate) \cite{sweeney2020future}.
It has become common now to decouple the estimation of the marginal probability density function of each variable and of the interdependence structure in the multivariate PWPF \cite{pinson2009probabilistic}.
In other words, univariate PWPF is usually recognized as the cornerstone of PWPF problems.

A classical approach for univariate PWPF relies on assumptions (often referred to as parametric approach) for the distribution of future wind power generation, the parameters of which are estimated via statistical and machine learning methods. For instance, the Gaussian, Beta, Generalized Logit-Normal, etc could be used \cite{morales2013integrating}. Although it is convenient to develop models based on such assumptions, the distribution of wind power at hand may not match the assumptions. This is primarily due to the wind power generation process, in other words, the nonlinear power curve that converts energy from the wind into electric power \cite{lange2005uncertainty}. Concretely, the characteristics of wind power generation distributions differ a lot depending on predicted weather conditions, as illustrated by \cite{pinson2010conditional} for instance. 
This has motivated many to look for distribution-free approaches, i.e., that do not rely on a specific assumption for the densities to model and communicate as forecasts.
Certainly the most popular distribution-free approach, also referred to as non-parametric, is quantile regression (QR) \cite{bremnes2006comparison}, which allows to relax the use of distributional assumptions for the case of univariate probabilistic forecasting. It has achieved great success in the Global Energy Forecasting Competition 2014~(GEFCom~2014) for instance, and has become a mainstream solution owing to its state-of-the-art performance and simplicity of use. However, it requires parallel models to be fitted for each quantile, which raises the cost of computation when the whole distribution is needed. In addition, it only provides discrete quantiles, which may lead to quantile crossing -- quantiles of the whole distribution are inconsistent.

Till now, parametric models with distributional assumption and QR are still the most effective methods \cite{sweeney2020future} with prominent characteristics. That is, parametric models characterize the whole distribution efficiently, whereas QR models are free of distributional assumptions.
For multivariate PWPF, the complex interdependence structures of multivariate distribution can be modeled using copula models \cite{tastu2015space}. By estimating the marginal probability density function (PDF) via non-parametric methods and modeling the complex interdependence structure, the copula method is allowed to model complicated multivariate distribution. However, the copula-based approach relies on strong assumptions regarding the probabilistic calibration of predicted marginals, while it often underestimate the strength of the dependence structure among the various variables. Eventually, it remains an open issue to develop an efficient, continuous and distribution-free probabilistic forecasting model that obtains whole distribution at once.

%\vspace{-1em}
\subsection{Related Works}

Univariate probabilistic forecasting usually translates to communicating quantile forecasts, prediction intervals (PIs), and predictive densities.
Quantile forecasts and PIs are specific characteristics of predictive densities, which are most often obtained by QR.
Based on this approach, several machine learning models such as neural network (NN) \cite{wan2016direct} and gradient boost machine \cite{LANDRY20161061} have been adopted to estimate conditional quantile functions.
It is then simple and effective to construct a PI with two corresponding quantile functions.
A $(1-\beta)\times 100\%$ PI can be constructed by the pair of quantiles $(\alpha,1-\beta+\alpha)$ where $\alpha \in (0,\beta)$. For instance, the pair $(\beta/2,1-\beta/2)$ is typically selected in the literature \cite{zhao2019adaptive,zhang2022optimal}.
However, both quantiles and PIs only provide partial information of probability densities, the applications of which can hardly cover power systems operation based on stochastic programming where the whole distribution of future wind power generation is often required.

As a result of this, it has been an active research topic to communicate densities in the PWPF community. Besides the aforementioned parametric approach, resampling and advanced density estimation techniques have been adopted, as reviewed in \cite{sweeney2020future,zhang2014review}.
The idea of resampling method lies in estimating the PDF of  empirical errors of point forecasts, which therefore makes the method distribution-free.
In order to issue conditional densities for the PWPF, fuzzy inference has been applied to classify the forecast conditions into several modes \cite{pinson2010conditional}.
But such finite classifications cannot continuously adapt to all forecast conditions.
Furthermore, the quality of estimated densities is strongly related to the performance of utilized point-forecast models.
The non-parametric density estimation method, namely kernel density estimation (KDE) has been popular among the PWPF community due to its universal approximation capability.
In particular, models based KDE usually deduce the density of a finite population selected by $k$-nearest neighbors \cite{zhang2016k}.
As with the resampling method, this method is still limited in modeling conditional densities, since the employed $k$-nearest neighbors operation is restricted in dealing with heterogeneous distributions.
That said, once $k$ is fixed, the KDE-based model cannot adaptively select the finite population.
In addition, the $k$-nearest neighbor operation suffers from the curse of dimension.
Recently, mixture density network (MDN) has been applied in PWPF, as it can model more complex distribution (compared to a Gaussian) through the comic combination of Gaussian distribution \cite{afrasiabi2020advanced}. 
But it would get stuck in mode collapse, i.e., the ultimate estimated distribution would collapse into a Gaussian distribution, and training instability \cite{makansi2019overcoming}.

Multivariate probabilistic forecasting often communicates \emph{scenarios} as forecasts, which are drawn from predictive densities.
The scenario generation procedure is based on probability integral transform~(PIT) and the interdependence structure \cite{pinson2009probabilistic}.
Concretely, one draws realizations from the estimated multivariate standard Gaussian distribution, and converts the realizations into scenarios of wind power generation via inverse PIT.
Besides, an emerging approach is to directly learn multivariate densities based on advanced generative models such as the generative adversarial network~(GAN) adopted in \cite{chen2018model}.
The GAN is composed of a generator and a discriminator, where the generator is responsible for generating scenarios at the operation stage.
Although it is computationally more efficient than the copula method, it suffers from notorious training instability caused by the game between the generator and discriminator at the training phase \cite{chu2019smoothness}.
Moreover, it only presents the applicability of GAN in generating scenarios, and as such is not focused on producing various forms of probabilistic forecasts e.g. predictive densities in univariate and multivariate setups.
Indeed, it has not even been assessed by proper statistical scores.
The most related work is \cite{dumas2021deep}, which compares the performance of several generative models, i.e., GAN, variational auto-encoder, and an integration-based normalizing flow~(NF).
But their primary focus is to compare the performance of deep-learning based generative models.
It is reported in \cite{dumas2021deep} that the performance of the integration-based NF is limited, let alone compared to state-of-the-art QR models.
Besides, they are unaware of the differences between affine NF (which is indeed is equivalent to parametric models with Gaussian distributional assumption) and integration-based NF models.
Therefore it leaves issues such as applicability of NF and the relationship between NF and existing models uncovered.

%%%%%%%%%%%%%%%% proposed method %%%%%%%%%%
\subsection{Proposed Method and Main Contributions}

As a basis for this work, we get inspiration from \cite{pinson2010conditional} and \cite{pinson2012very}, which relied on the idea of transforming samples of bounded stochastic process at hand to make them more suitable to be modeled by a Gaussian (or multivariate Gaussian) variable.
Besides, parametric models always serve as good candidates for estimating the underlying distributions of wind power generation \cite{wen2022sparse}. Thus, it is appealing to set a parametric model to learn a base distribution, and transform the base distribution to the desired distribution (in the view of the underlying distribution of wind power generation) with an affordable cost.
Indeed, it is allowed by the \emph{conservation of probability measure}~\cite{stein2003princeton}, which translates into saying that one can transform a variable that follows an arbitrary distribution into a variable that follows a desired distribution with the assistance of bijective mapping~(transform).
Here, instead of using a manually designed transform, we implement such transforms via the NF~\cite{rezende2015variational,papamakarios2021normalizing}.
An NF framework is composed of a base distribution and a sequence of trainable bijective mappings.
Both the shape parameters of base distribution and bijective mappings are modeled by neural networks~(NNs).
Besides, such transforms ought to be non-affine so that the model can flexibly characterize the wind power distribution under different conditions.

Concretely, we establish a distribution-free PWPF model based on a combination of a parametric model with Gaussian distributional assumption and a conditional auto-regressive NF \cite{papamakarios2017masked}, which is applicable to both univariate and multivariate PWPF applications.
Unlike copula models where the marginal PDF and interdependence structure are modeled separately, here the joint probability density is derived through the chain rule of probability, i.e., the product of conditional probability densities. In particular, such conditional probability densities are also dependent on input features.
The base Gaussian distribution are estimated by the parametric model, whose realizations are then mapped into those of the desired distribution via a spline-based NF~\cite{durkan2019neural}.
By using the non-affine characteristics of spline-based NF, the model is allowed to characterize the predicted distribution of wind power generation more flexibly.
The spline operates in an elementwise manner, i.e., the mapping for each dimension is specified by the outputs of an NN that takes contextual features and the values of previous dimension as inputs.
All the parameters are estimated simultaneously based on the maximum likelihood.
Case studies validate the effectiveness of the proposed model, which achieves state-of-the-art.

The main contributions of the paper are: (\romannumeral1) The proposal of a distribution-free PWPF model, which suffices to handle the bounded characteristics of wind power by using the power of a parametric model with Gaussian distributional assumption and non-affine transforms.
(\romannumeral2) The demonstration of its applicability to model the whole predictive distribution, which avoids the quantile crossing issue in the univariate PWPF and still presents competitive performance that is comparable to state-of-the-art QR models.
(\romannumeral3) A new perspective for conditional PDF estimation for PWPF based on the function theory, which offers complimentary understanding to merits and caveats of distribution-free approaches versus parametric approaches.

The remainder of this paper is organized as follows.  In section \uppercase\expandafter{\romannumeral2}, the problem formulation and methodological components of normalizing flows are introduced. Our approach to their application to univariate and multivariate wind power probabilistic forecasting is described in section \uppercase\expandafter{\romannumeral3}. Section \uppercase\expandafter{\romannumeral4} summarizes data sources and experiment implementation. The results obtained are presented in Section \uppercase\expandafter{\romannumeral5}, where the performance comparison with existing models is discussed. Section \uppercase\expandafter{\romannumeral6} concludes this paper.

\begin{figure}[!t]
\centering
\includegraphics[width=3.4in]{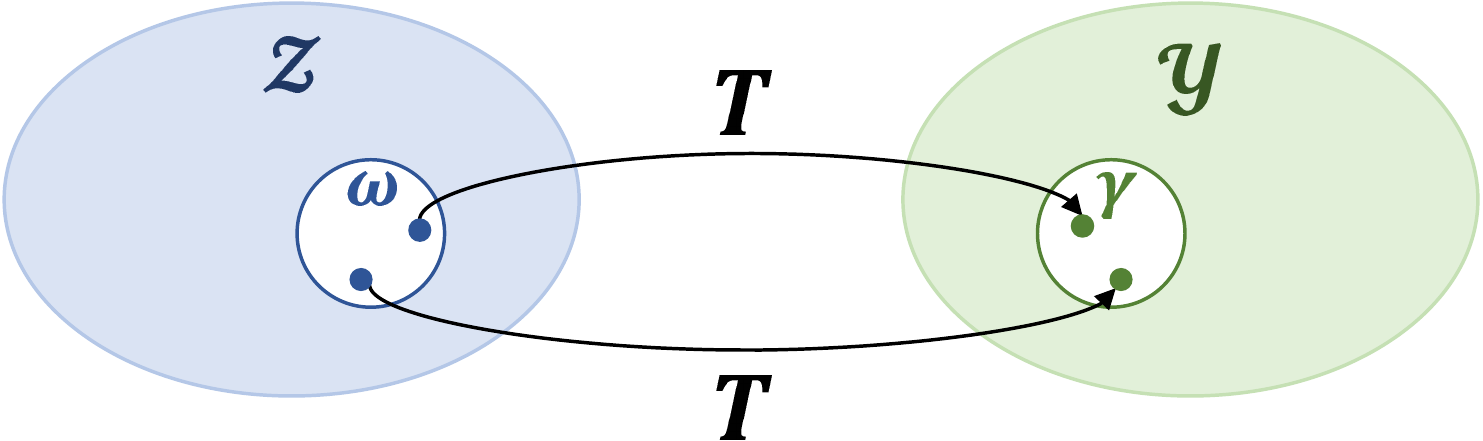}
\caption{Illustration of transform.}
\label{transform}
\vspace{-1em}
\end{figure}

\section{Methodological Components}

\subsection{Preliminaries}

The most important base property to consider for normalizing flows is the concept of conservation of probability measure.

\textbf{\emph{Definition~1} (Conservation of Probability Measure)} : Denote the PDF defined on $\mathcal{Z} \subseteq \mathbb{R}^d$ as $f_Z(\boldsymbol z): \mathcal{Z}\rightarrow [0,+\infty) $, the PDF defined on $\mathcal{Y}\subseteq \mathbb{R}^d$ as $f_Y(\boldsymbol y): \mathcal{Y}\rightarrow [0,+\infty) $, and an invertible transform as $T: \mathcal{Z}\rightarrow \mathcal{Y} $.
For any subset $\omega \subseteq \mathcal{Z}$, we have
\begin{equation}
    \int_{\boldsymbol z \in \omega}f_Z(\boldsymbol z)d \boldsymbol z=\int_{\boldsymbol y \in \gamma}f_Y(\boldsymbol y)d \boldsymbol y.
\end{equation}
where $\gamma=\{T(\boldsymbol z)|\boldsymbol z \in \omega \}$, as illustrated in Fig.~\ref{transform}. By utilizing the change of variable, $\boldsymbol z = T^{-1}(\boldsymbol y)$, we convert the formula into
\begin{align*}
    \int_{\boldsymbol y \in \gamma}f_Z(T^{-1}(\boldsymbol y))| \det J_{T^{-1}}(\boldsymbol y)|d \boldsymbol y=\int_{\boldsymbol y \in \gamma}f_Y(\boldsymbol y)d \boldsymbol y,
\end{align*}
where $J_{T^{-1}}(\boldsymbol y)$ denotes the Jacobian matrix s.t. 
\begin{align*}
    J_{T^{-1}}(\boldsymbol y)_{i.j}=\frac{\partial y_i}{\partial z_j}.
\end{align*}
As it holds for any subset $\gamma \subseteq \mathcal{Y}$, we have
\begin{equation}
    f_Y(\boldsymbol y)=f_Z(T^{-1}(\boldsymbol y))| \det J_{T^{-1}}(\boldsymbol y)|.
\end{equation}

\subsection{Problem Formulation}

Consider we have $p$ wind farms whose generation is driven by a multivariate stochastic process.
For wind farm $i$, let $y_{i,t}$ denote the generation value at time $t$, which is a realization of the corresponding random variable $Y_{i,t}$.
Then, let $f_{Y_{i,t}}(y)$ and $F_{Y_{i,t}}(y)$ respectively denote the PDF and cumulative distribution function (CDF) of $Y_{i,t}$.
The univariate PWPF boils down to estimating the PDF of $Y_{i,t+H}$, i.e., $\hat{f}_{Y_{i,t+H}|t}$, given information $\Omega_{i,t}$ up to $t$ via a model $\mathcal{M}$, i.e.,
\begin{equation}
    \hat{f}_{Y_{i,t+H}|t}=f_{Y_{i,t+H}|t}(y|\Omega_{i,t};\mathcal{M},\hat{\Theta}),
\end{equation}
where $H$ is the forecasting horizon, and $\hat{\Theta}$ represents the estimation of real parameters $\Theta$.
Certainly, information from nearby wind farms could be used to improve the forecasts, if available \cite{cavalcante2017lasso}.
The information may contain previous wind power generation values, i.e., $\{ y_{i,t-l},\cdots,y_{i,t-1},y_{i,t} \}$, and some exogenous features such as numerical weather predictions (NWPs).
Accordingly, one can also obtain the CDF of $Y_{i,t+H}$ by integrating $\hat{f}_{Y_{i,t+H}|t}$, namely $\hat{F}_{Y_{i,t+H}|t}$, the inverse function of which specifies quantiles.
For instance, the predicted $\alpha$-th quantile $\hat{q}^{(\alpha)}_{t+H|t}$ is given by
\begin{equation}
    \hat{q}^{(\alpha)}_{t+H|t}=\hat{F}^{-1}_{Y_{i,t+H}|t}(\alpha).
\end{equation}
A PI with nominal level $(1-\beta)\times 100\%$ can be formed by two quantiles, $\hat{q}^{(\beta/2)}_{t+H|t}$ and $\hat{q}^{(1-\beta/2)}_{t+H|t}$, i.e.,
\begin{equation}
    \left[\hat{q}^{(\beta/2)}_{t+H|t},\ \hat{q}^{(1-\beta/2)}_{t+H|t}\right].
\end{equation}

Indeed, multivariate PWPF aims at communicating the joint probability distribution of a collection of future random variables.
For instance, the multivariate PWPF may communicate the joint probability distribution of random variables at several future time, i.e., $Y_{i,t+1},\cdots,Y_{i,t+H}$, which is expressed as
\begin{equation}
    \hat{f}_{Y_{i,t+1},\cdots,Y_{i,t+H}|t}=
    f_{Y_{i,t+1},\cdots,Y_{i,t+H}|t}(\boldsymbol y|\Omega_{i,t};\mathcal{M},\hat{\Theta}), 
\end{equation}
and that at several sites, i.e.,
\begin{equation}
    \begin{split}
        &\hat{f}_{Y_{1,t+H},\cdots,Y_{p,t+H}|t}=\\
        &f_{Y_{1,t+H},\cdots,Y_{p,t+H}|t}(\boldsymbol y|\Omega_{1,t},\cdots,\Omega_{p,t};\mathcal{M},\hat{\Theta}).
    \end{split}
\end{equation}
In multivariate PWPF, one often draws several realizations as scenarios from the estimated distribution.
For instance, one can draw realizations from $\hat{f}_{Y_{1,t+H},\cdots,Y_{p,t+H}|t}$, which are denoted as $\Tilde{y}^{(s)}_{1,t+1},\cdots,\Tilde{y}^{(s)}_{p,t+H}$, i.e.,
\begin{equation}
    \Tilde{y}^{(s)}_{1,t+1},\cdots,\Tilde{y}^{(s)}_{p,t+H}\sim \hat{f}_{Y_{1,t+H},\cdots,Y_{p,t+H}|t}.
\end{equation}
Without loss of generality, we write the future random variable as $\boldsymbol Y_t$ (which may be univariate or multivariate), and its realization as $\boldsymbol y_t$.
The information is denoted as $\Omega_t$, whose realization is $\boldsymbol x_t$.
In this paper, we refer to $\boldsymbol x_t$ as contextual features, to make them distinguished from the inputs of NF. 
Hence, the cornerstone of PWPF can be written in a compact form, i.e.,
\begin{equation}
    \hat{f}_{\boldsymbol Y_t|t}(\boldsymbol y|\boldsymbol x_t)=f_{\boldsymbol Y_t|t}(\boldsymbol y|\boldsymbol x_t;\mathcal{M},\hat{\Theta}).
\end{equation}
In this paper, we assume that $\Theta$ does not change with time, which therefore can be estimated from training datasets $\boldsymbol X_{train}$ and $\boldsymbol Y_{train}$.
It can be also considered in an online setting, where parameters vary with time.
With the estimated model at hand, to issue a forecast at time $t$, it is only required to feed $\boldsymbol x_t$ into the model and yield results as described in (9).

\begin{figure}[!t]
\centering
\includegraphics[width=2.5in]{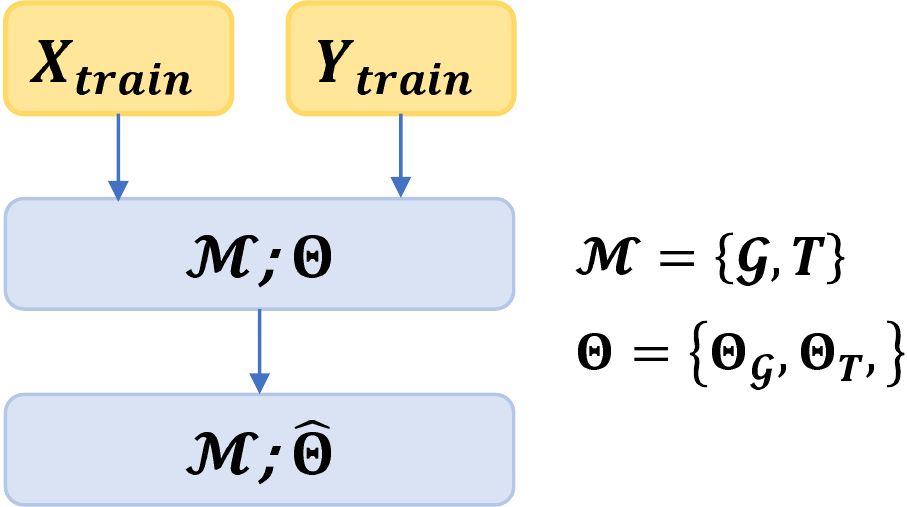}
\caption{Illustration of model estimation stage.}
\label{model estimation}
\end{figure} 

The classic parametric approach usually sets $\mathcal{M}$ as a model with distributional assumption, such as Gaussian and Logit-normal, whereas $\hat{\Theta}$ denotes the parameters of a function that maps contextual features to the shape parameters of distribution.
With the conservation of probability measure, we consider an intermediate random variable $\boldsymbol Z_t$ that follows a specific distribution $f_{\boldsymbol Z_t}(\boldsymbol z)$, whose realization is denoted as $\boldsymbol z_t$. Let $T$ map $\boldsymbol z_t$ into $\boldsymbol y_t$, i.e.,
\begin{equation}
    \boldsymbol y_t = T(\boldsymbol z_t;\hat{\Theta}_T),
\end{equation}
where $\hat{\Theta}_T$ denotes the estimation of parameters of transform $T$ (whose real parameters are denoted as $\Theta_T$).
Now we consider to model the distribution of $\boldsymbol Z_t$ via a parametric model $\mathcal{G}$, whose parameters are denoted as $\Theta_{\mathcal{G}}$. The estimation of $\Theta_{\mathcal{G}}$ is denoted as $\hat{\Theta}_{\mathcal{G}}$.
Then, the model $\mathcal{M}$ consists of $\mathcal{G}$ and $T$, i.e., $\mathcal{M}=\{ \mathcal{G}, T \}$, whose parameters are $\Theta=\{\Theta_{\mathcal{G}}, \Theta_T \}$.
The conceptual framework of training stage is shown in Fig.~\ref{model estimation}.
In other words, by learning $\mathcal{G}$ and $T$, we can estimate the model $\mathcal{M}$.

\begin{figure}[!t]
\centering
\includegraphics[width=3.25in]{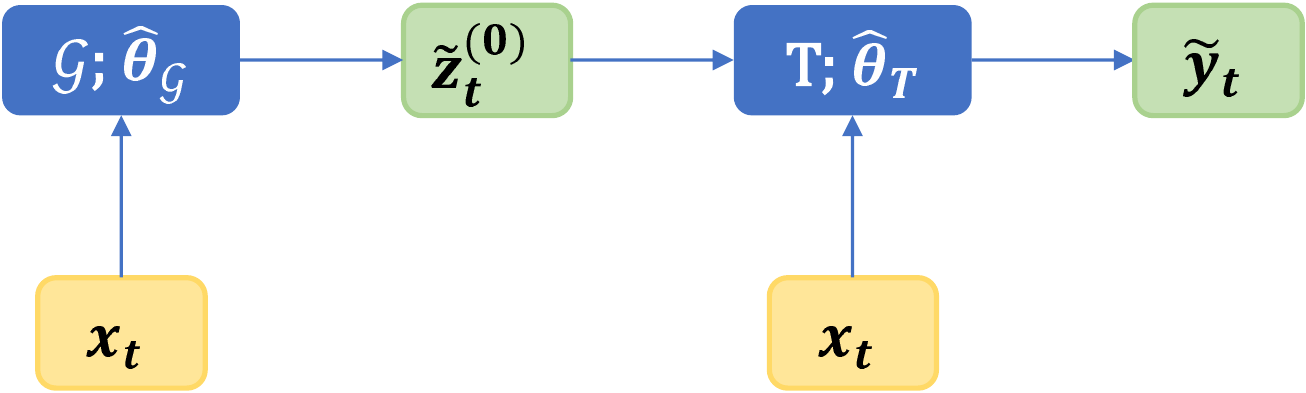}
\caption{Illustration of operational forecasting stage.}
\label{operational}
\end{figure} 

\subsection{Flow Model for Forecasting}
Here we implement the conceptual model $T$ via normalizing flows.
Generally, the transform $T$ in an NF consists of a series of invertible functions $T_1, T_2, \dots, T_K$~\cite{papamakarios2021normalizing}, i.e.,
\begin{equation}
    T=T_1 \circ T_2 \circ \cdots \circ T_K,
\end{equation}
where $\circ$ denotes the symbol of composition. 
For each $T_k$, we denote its input as $\boldsymbol z^{(k-1)}_t$, which is the realization of the random variable $\boldsymbol Z^{(k-1)}_t$.
Accordingly, its output is denoted as $\boldsymbol z^{(k)}_t$, which is the realization of the random variable $\boldsymbol Z^{(k)}_t$.
Particularly, $\boldsymbol Z^{(0)}_t$ follows the base distribution specified by $ \mathcal{G}$, whereas $\boldsymbol Z^{(K)}_t$ is $\boldsymbol Y_t$.
For simplicity of notations, we drop subscript of density function in what follows.

Two significant calculation passes in NF models are forward and inverse passes.
Such computation between $\boldsymbol z^{(k)}_t$ and $\boldsymbol z^{(k-1)}_t$ for instance is respectively described as
\begin{align*}
    \boldsymbol z^{(k)}_t = T_k(\boldsymbol z^{(k-1)}_t;\hat{\Theta}_{T_k}),\quad \boldsymbol z^{(k-1)}_t = T_k^{-1}(\boldsymbol z^{(k)}_t;\hat{\Theta}_{T_k}),
\end{align*}
where $\hat{\Theta}_{T_k}$ represents the estimated parameters of $T_k$.
In particular, we obtain $f(\boldsymbol z^{(k)}_t|\boldsymbol x_t)$ through $f(\boldsymbol z^{(k-1)}_t|\boldsymbol x_t)$ and the mapping $T_k$, which is bijective in $\boldsymbol z^{(k-1)}_t$ as well as $\boldsymbol z^{(k)}_t$ and parameterized by $\boldsymbol x_t$ \cite{winkler2019learning}.
We have
\begin{equation}
\begin{split}
    f&(\boldsymbol z^{(k)}_t|\boldsymbol x_t)=
    f(\boldsymbol z^{(k-1)}_t|\boldsymbol x_t)|\frac{\partial \boldsymbol z^{(k-1)}_t}{\partial \boldsymbol z^{(k)}_t}|\\
    &=f(T_k^{-1}(\boldsymbol z^{(k)}_t;\hat{\Theta}_{T_k},\boldsymbol x_t)|\boldsymbol x_t)|\det J_{T_k}(\boldsymbol z^{(k-1)}_t)|.
\end{split}
\end{equation}
Consequently, the forward and inverse passes in CNF are expressed as
\begin{equation}
    \boldsymbol z^{(k)}_t = T_k(\boldsymbol z^{(k-1)}_t;\hat{\Theta}_{T_k},\boldsymbol x_t),\quad \boldsymbol z^{(k-1)}_t = T_k^{-1}(\boldsymbol z^{(k)}_t;\hat{\Theta}_{T_k},\boldsymbol x_t).
\end{equation}
With the sequential transforms, we have
\begin{align*}
    \boldsymbol y_t = T(\boldsymbol z^{(0)}_t;\hat{\Theta}_T,\boldsymbol x_t),\quad \boldsymbol z^{(0)}_t = T^{-1}(\boldsymbol y_t;\hat{\Theta}_T,\boldsymbol x_t).
\end{align*}
where $\hat{\Theta}_T$ denotes the parameters of $T$, which is a collection of $\hat{\Theta}_{T_k}$, i.e., $\hat{\Theta}_T=\{\hat{\Theta}_{T_1},\cdots,\hat{\Theta}_{T_K} \}$.

The Jacobian determinant is computed by
\begin{align*}
\begin{split}
    \log |\det J_T(\boldsymbol z^{(0)}_t)|&=
    \log |\prod_{k=1}^K \det J_{T_k}(\boldsymbol z^{(k-1)}_t)|\\
    &=\sum_{k=1}^K \log |\det J_{T_k}(\boldsymbol z^{(k-1)}_t)|.
\end{split}
\end{align*}
Ultimately, we build the connection between the PDF of $\boldsymbol z^{(0)}_t$ and that of $\boldsymbol y_t$, i.e.,
\begin{align*}
    \log f(\boldsymbol y_t)= \log f(\boldsymbol z^{(0)}_t)+\sum_{k=1}^K \log |\det J_{T_k}(\boldsymbol z^{(k-1)}_t)|.
\end{align*}
Such $T_k$ in the NF model is implemented via NNs, and is required to be invertible and have a tractable Jacobian determinant.

The introduced CNF model is trained based on maximum likelihood.
As we assume that parameters $\Theta$ will not change with time, we can estimate them from training dataset  $\boldsymbol Y_{train}=[\boldsymbol y_1,\boldsymbol y_2,\cdots,\boldsymbol y_N]^\top $ and $\boldsymbol X_{train}=[\boldsymbol x_1,\boldsymbol x_2,\cdots,\boldsymbol x_N ]^\top$.
The loss function is defined as
\begin{equation}
\begin{split}
    &\mathcal{L}=-\frac{1}{N}\sum_{n=1}^N \log f(\boldsymbol y_n|\boldsymbol x_n)\\
    &=-\frac{1}{N}\sum_{n=1}^N[\log f(T^{-1}(\boldsymbol y_n;\boldsymbol x_n))+
    \log |\det J_T(T^{-1}(\boldsymbol y_n;\boldsymbol x_n))|].
\end{split}
\end{equation}
At the training stage, we estimate $\hat{\Theta}$ by minimizing the loss function $\mathcal{L}$.
To issue a forecast at time $t$, we feed $\boldsymbol x_t$ into the base model and all transforms, which is illustrated in Fig.~\ref{operational}.
Then, we derive the density of $\boldsymbol z_t^{(0)}$, i.e.,
\begin{align*}
    \hat{f}(\boldsymbol z_t^{(0)}|\boldsymbol x_t;\mathcal{G},\hat{\Theta}_{\mathcal{G}}).
\end{align*}
Based on it, we could draw $L$ realizations:
\begin{equation}
    \Tilde{\boldsymbol z}_t^{(0),1},\cdots,\Tilde{\boldsymbol z}_t^{(0),L} \sim \hat{f}(\boldsymbol z_t^{(0)}|\boldsymbol x_t;\mathcal{G},\hat{\Theta}_{\mathcal{G}}).
\end{equation}
By transforming each realization $\Tilde{\boldsymbol z}_t^{(0),i}$ via $T$, i.e.,
\begin{equation}
    \Tilde{\boldsymbol y}_t^i=T(\Tilde{\boldsymbol z}_t^{(0),i};\boldsymbol x_t, \hat{\Theta}_T),
\end{equation}
we can obtain $L$ realizations of $\hat{f}(\boldsymbol y_t|\boldsymbol x_t;\mathcal{M},\hat{\Theta})$, namely $\Tilde{\boldsymbol y}_t^1,\cdots,\Tilde{\boldsymbol y}_t^L$.
In particular, we can obtain the $\alpha$-th quantile of $\hat{f}(\boldsymbol z_t^{(0)}|\boldsymbol x_t;\mathcal{G},\hat{\Theta}_{\mathcal{G}})$, which is denoted as $\hat{q}_{\mathcal{G}}^{(\alpha)}$, and then transform it via $T$ to obtain the quantile of $\hat{f}(\boldsymbol y_t|\boldsymbol x_t;\mathcal{M},\hat{\Theta})$, i.e.,
\begin{equation}
    \hat{q}_{\mathcal{M}}^{(\alpha)}=T(\hat{q}_{\mathcal{G}}^{(\alpha)};\boldsymbol x_t,\hat{\Theta}_T)
\end{equation}

\subsection{Relationship with Classic Methods}

Here we discuss the relationship between this method and classic methods.
In what follows, we assume the base distribution as a standard normal distribution, i.e., $\boldsymbol z^{(0)}_t\sim \mathcal{N}(\boldsymbol 0,\boldsymbol I)$.

\subsubsection{Gaussian Distribution}
Models with Gaussian distributional assumption \cite{Khosravi2011,kou2013sparse}
are described as
\begin{align*}
    \boldsymbol y_t \sim \mathcal{N}(\boldsymbol \mu_t(\boldsymbol x_t),\Sigma_t(\boldsymbol x_t)),
\end{align*}
where $\boldsymbol \mu_t(\boldsymbol x_t)$ and $\Sigma_t(\boldsymbol x_t)$ are the corresponding shape parameters, which are specified by $\boldsymbol x_t$ and estimated via statistical learning.
They can be translated into setting the transform $T$ as the composition of affine transforms. That is,
\begin{align*}
    \boldsymbol y_t = T(\boldsymbol z^{(0)}_t;\boldsymbol x_t) = \boldsymbol A_t(\boldsymbol x_t) \boldsymbol z^{(0)}_t+\boldsymbol b_t(\boldsymbol x_t),
\end{align*}
where $\boldsymbol A_t(\boldsymbol x_t)$ and $\boldsymbol b_t(\boldsymbol x_t)$ are the corresponding matrix and vector specified by $\boldsymbol x_t$.
Then the problem boils down to estimating $\boldsymbol A_t(\boldsymbol x_t)$ and $\boldsymbol b_t(\boldsymbol x_t)$ from data.
As affine transforms cannot change the family of distributions, $\boldsymbol y_t$ still obeys Gaussian distribution,

\subsubsection{Logit-Normal Distribution}

The logit-normal distribution \cite{pinson2012very} can be derived by applying a logit-normal transform to a Gaussian distribution, i.e.,
\begin{align*}
    \boldsymbol y_t \sim L(\boldsymbol \mu(\boldsymbol x_t),\Sigma_t(\boldsymbol x_t)).
\end{align*}
It can be interpreted as setting the transform $T$ in a normalizing flow as a combination of affine transforms and a sigmoid transform.
Using the affine transforms, we derive
\begin{align*}
    \boldsymbol z^{(K-1)}_t \sim \mathcal{N}(\boldsymbol \mu_t(\boldsymbol x_t),\Sigma_t(\boldsymbol x_t)),
\end{align*}
where $\boldsymbol \mu_t(\boldsymbol x_t)$ and $\Sigma_t(\boldsymbol x_t)$ are specified by $\boldsymbol x_t$.
Then the logit-normal transform operates element-wise on $\boldsymbol z^{(K-1)}_t$, i.e.,
\begin{align*}
    y_{t,i} = \frac{\exp(z^{(K-1)}_{t,i})}{ 1+\exp( z^{(K-1)}_{t,i})},
\end{align*}
where $y_{t,i}$ and $z_{t,i}^{(K-1)}$ respectively represent the $i$-th element of $\boldsymbol y_t$ and $\boldsymbol z_t^{(K-1)}$.

\subsubsection{Mixture Density Network}
Mixture density network is a popular model that outputs the parameters of Gaussian mixture models. It is described as
\begin{align*}
    f(\boldsymbol y_t|\boldsymbol x_t)=\sum\pi_i(\boldsymbol x_t)f(\boldsymbol y_t;\boldsymbol \mu_i(\boldsymbol x_t),\boldsymbol \Sigma_i(\boldsymbol x_t)),
\end{align*} where $\sum \pi_i(\boldsymbol x) = 1$.
Models based on mixture density networks can be regarded as setting $T$ as a conic combination of affine transforms.
That is, 
\begin{align*}
    \boldsymbol y_t = T(\boldsymbol z^{(0)}_t;\boldsymbol x_t)=\sum\pi_i(\boldsymbol x_t) T_i(\boldsymbol z^{(0)}_t;\boldsymbol x_t),
\end{align*}
where $T_i$ operates as
\begin{align*}
    T_i(\boldsymbol z^{(0)}_t; \boldsymbol x_t)=\boldsymbol A^i_t(\boldsymbol x_t) \boldsymbol z^{(0)}_t + \boldsymbol b^i_t(\boldsymbol x_t),
\end{align*}
where $\boldsymbol A^i_t$ and $\boldsymbol b^i_t$ are parameters, specified by $\boldsymbol x_t$.

\subsubsection{Gaussian Copula}
The model based on Gaussian Copula \cite{pinson2009probabilistic} is an instance of NF, which is specified by an element-wise monotone function $g$ and a correlation matrix $\Sigma_t$ specified by $\boldsymbol x_t$.
That is, 
\begin{align*}
    \boldsymbol z^{(K-1)}_t= \boldsymbol A_t(\boldsymbol x_t) \boldsymbol z^{(0)}_t \sim \mathcal{N}(\boldsymbol 0, \Sigma_t(\boldsymbol x_t)),
\end{align*}
\begin{align*}
    y_{t,i} = g(z^{(K-1)}_{t,i};\boldsymbol x_t).
\end{align*}

Indeed, any desired distribution can be obtained by transforming a Gaussian distribution through a specific mapping. 
Such mapping proceeds each value in the domain in the same manner, such as the aforementioned Logit-Normal transform.
%It implies that characteristics of the derived wind power distributions remain the same for different wind conditions.
Therefore, the mapping is required to be specified by conditional information, so that the derived distribution is allowed to adapt to different wind conditions.
Although the conic combination enables deriving more complex distributions compared to Gaussian distributions, it is restricted by the number of mixing components.
With regard to the Gaussian copula model, it is developed for multivariate modeling. 
By modeling the well-calibrated marginal PDF and correlation structure, one can yield the the ultimate joint probability density in a distribution-free way. 
However, as mentioned above, it highly relies on the estimation of marginals and tends to underestimate the covariance structure, which often impedes its performance.

\begin{figure}[!t]
\centering
\includegraphics[width=3.4in]{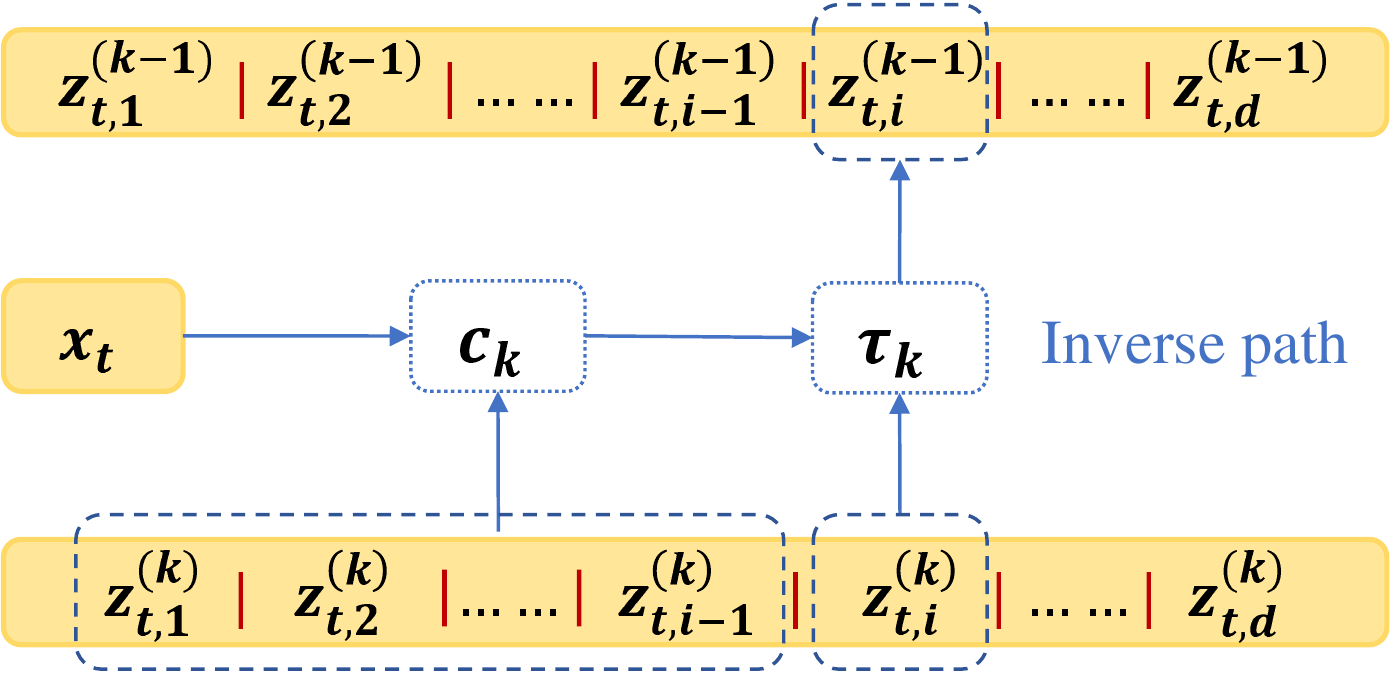}
\caption{Illustration of inverse path in the $k$-th transform.}
\label{inverse}
\end{figure}

\begin{figure}[!t]
\centering
\includegraphics[width=3.4in]{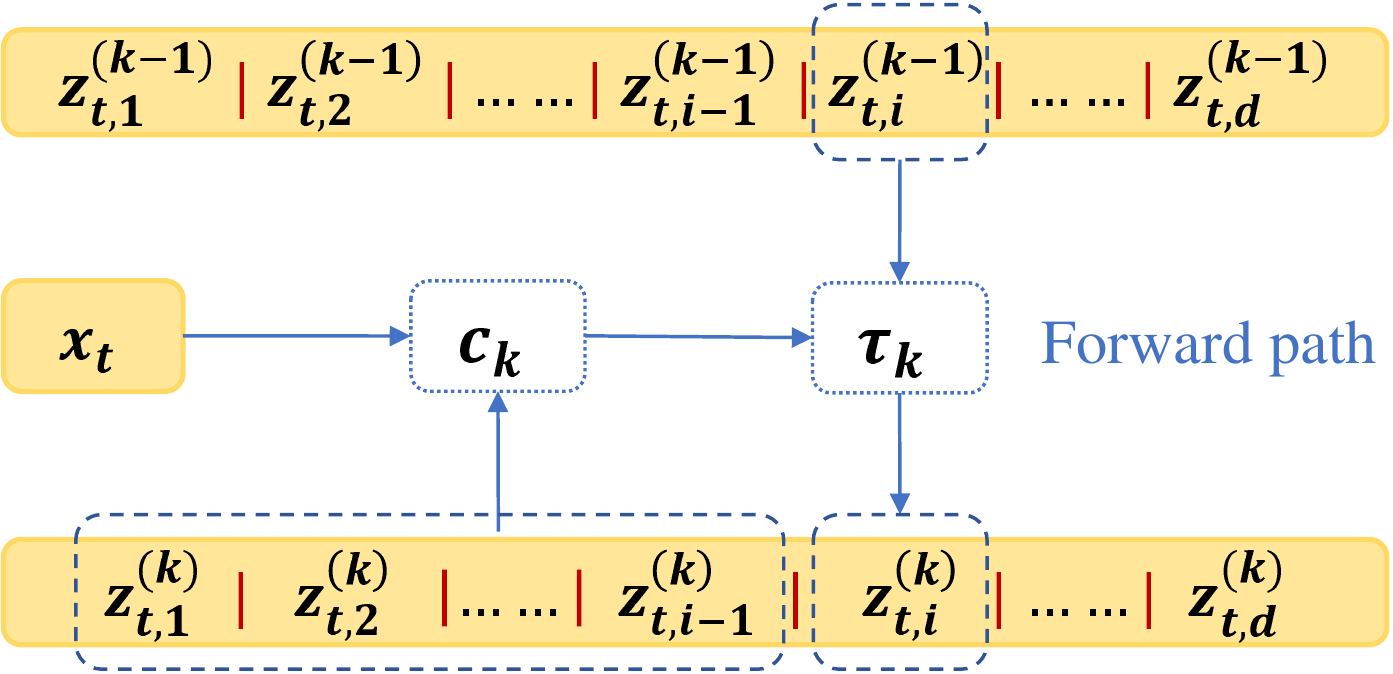}
\caption{Illustration of forward path in the $k$-th transform.}
\label{forward}
\end{figure}

\section{Forecasting Applications}

The basic approach for conditional normalizing flows described in the above can readily be used for forecasting applications, in both univariate and multivariate settings. We choose a Gaussian distribution as the base distribution whose shape parameters are learned by an NN and adopt a non-affine flow to obtain a piece-wise non-Gaussian distribution. 
Let $\hat{\boldsymbol \mu}_t, \hat{\Sigma}_t$ denote the estimated shape parameters of base distribution, which are determined by a function of $\boldsymbol x_t$, namely $\phi(\boldsymbol x_t;\hat{\Theta}_{\mathcal{G}})$. In other words, with the Gaussian distributional assumption, the model $\mathcal{G}$ described in Section~II-B reduces to the function $\phi(\boldsymbol x_t;\hat{\Theta}_{\mathcal{G}})$.
It is described as
\begin{equation}
    \hat{\boldsymbol \mu}_t, \hat{\Sigma}_t = \phi(\boldsymbol x_t;\hat{\Theta}_{\mathcal{G}}).
\end{equation}

\subsection{Probabilistic Forecasting Applications}
\subsubsection{Univariate Probabilistic Forecasting}
In the univariate case, each intermediary variable (for instance the $k$-th intermediary variable) and shape parameters of the Gaussian distribution are scalars, which are rewritten as $z^{(k)}_t$, $\mu_t$, and $\sigma_t$.
The estimated shape parameters of base distribution $\hat{\mu}_t$, $\hat{\sigma}_t$ are derived via 
\begin{equation}
    \hat{\mu}_t, \hat{\sigma}_t = \phi(\boldsymbol x_t;\hat{\Theta}_{\mathcal{G}}).
\end{equation}
$T_k$ is a univariate function that operates as
\begin{equation}
    z^{(k)}_t=T_k(z^{(k-1)}_t;\boldsymbol x_t,\hat{\Theta}_T).
\end{equation}

\subsubsection{Multivariate Probabilistic Forecasting}
The most relevant computation to consider for multivariate forecasting is the transform described in (13).
Here, let us consider a function $\tau_k$ in the transform $T_k$ that operates elementwise and relies on previous dimensions and contextual information.
Take the computation of $i$-th dimension as an example, i.e., the forward path and inverse path between $z^{(k-1)}_{t,i}$ and $z^{(k)}_{t,i}$.
In the inverse path, $\tau^{-1}_k$ maps $z^{(k)}_{t,i}$ into $z^{(k-1)}_{t,i}$ via
\begin{equation}
    z^{(k-1)}_{t,i}=\tau^{-1}_k(z^{(k)}_{t,i};c_k(\boldsymbol z^{(k)}_{t,1:i-1},\boldsymbol x_t;\hat{\theta}_{c_k}),\hat{\theta}_{\tau_k}),
\end{equation}
where $\boldsymbol z^{(k)}_{t,1:i-1}$ represents $[z^{(k)}_{t,1},\cdots,z^{(k)}_{t,i-1}]^\top$, $\hat{\theta}_{\tau_k}$ represents the parameters of $\tau_k$, and $c(\boldsymbol z^{(k)}_{t,1:i-1},\boldsymbol x_t;\hat{\theta}_{c_k})$ is a function that outputs conditionals.
In other words, $\hat{\Theta}_{T_k}$ contains $\hat{\theta}_{\tau_k}$ and $\hat{\theta}_{c_k}$.
The forward path is described as
\begin{equation}
    z^{(k)}_{t,i}=\tau_k(z^{(k-1)}_{t,i};c_k(\boldsymbol z^{(k)}_{t,1:i-1},\boldsymbol x_t;\hat{\theta}_{c_k}),\hat{\theta}_{\tau_k}).
\end{equation}
Using the terminology of \cite{pmlr-v80-huang18d}, $c_k(\cdot)$ and $\tau_k(\cdot)$ are respectively referred to as the conditioner and transformer.
Illustration of such calculation procedure is shown in Fig.~\ref{inverse} and Fig.~\ref{forward}.

\emph{Remark 1}: With the chain rule of probability, we decompose the joint probability density $f(\boldsymbol z^{(k)}_t|\boldsymbol x_t)$ into a product of conditional probability densities, i.e.,
\begin{align*}
    f(\boldsymbol z^{(k)}_t|\boldsymbol x_t)=\prod_{i=1}^d f(z^{(k)}_{t,i}|\boldsymbol z^{(k)}_{t,1:i-1},\boldsymbol x_t).
\end{align*}
As shown in Section~II-C, the training stage is relied on the inverse path and the computation of likelihood.
Indeed, the inverse path described in (21) is associated with $f(z^{(k)}_{t,i}|\boldsymbol z^{(k)}_{t,1:i-1},\boldsymbol x_t)$, which translates into saying that the computation of likelihood will preserve the conditional structure of multivariate distribution.
The forward path can be translated into sampling $\boldsymbol z^{(k-1)}_t$ from $f(\boldsymbol z^{(k-1)}_t|\boldsymbol x_t)$ and computing via (22), which can be also regarded as sampling $z^{(k)}_{t,i}$ from $f(z^{(k)}_{t,i}|\boldsymbol z^{(k)}_{t,1:i-1},\boldsymbol x_t)$.

\emph{Remark 2}: The univariate probabilistic forecasting can be interpreted as a special case of multivariate probabilistic forecasting.
As with (22), we rewrite (20) as
\begin{align*}
    z_t^{(k)}=\tau_k(z_t^{(k-1)};c_k(\boldsymbol x_t;\hat{\theta}_{c_k}),\hat{\theta}_{\tau_k}).
\end{align*}

\begin{table*}[!t]
\renewcommand{\arraystretch}{1.25}
\caption{Case study settings.}
\label{table_case}
\centering
\begin{tabular}{m{1cm}|m{2cm}<{\centering}|m{3.5cm}<{\centering}|m{1.5cm}<{\centering}|m{3cm}<{\centering}|m{3cm}<{\centering}}
\hline
 & Type of variable & Input feature & Forecasting horizon & Type of interdependence  & Dataset \cr
\hline
Case 1 & univariate& NWP& 24&  none& GEFCom 2014 \cr
\hline
Case 2 & univariate& previous values of length 6& 1&  none& NREL, France wind farm \cr
\hline
Case 3 & multivariate& previous values of length 6& 6&  temporal interdependence& France wind farm \cr
\hline
Case 4 & multivariate& previous values of length 6& 1&  spatial interdependence& NREL \cr
\hline
\end{tabular}
\vspace{-1em}
\end{table*}

\begin{comment}
\begin{table*}[!t]
\renewcommand{\arraystretch}{1.25}
\caption{Dataset description.}
\label{table_data}
\centering
\begin{tabular}{m{3.25cm}|m{8cm}|m{1.75cm}<{\centering}|m{1.5cm}<{\centering}}
\hline
Dataset & Description & Resolution & Samples  \cr
\hline
GEFCom 2014 & NWP that contains wind speed and direction at two altitudes respectively, as well as corresponding wind power values& 1-h& 16800  \cr
\hline
France wind farm & Time series of wind power& 10-min& 52355 \cr
\hline
NREL & Time series of wind power& 15-min& 35040 \cr
\hline
\end{tabular}
\end{table*}
\end{comment}

\subsection{Base Distribution}
The function $\phi(\cdot)$ described in (18) and (19) is implemented by an NN of $N_{\phi}$ layers.
Denote the outputs, weights, and bias of the $l$-th layer respectively as $\boldsymbol h^{\phi,l}_{t}$, $\boldsymbol W^{\phi,l}$, and $\boldsymbol b^{\phi,l}$.
The $l$-th layer operates as
\begin{equation}
    \boldsymbol h^{\phi,l}_{t} = \boldsymbol W^{\phi,l} \boldsymbol h^{\phi,l-1}_{t}+\boldsymbol b^{\phi,l}.
\end{equation}
Specially, $\boldsymbol h^{\phi,0}_{t}=\boldsymbol x_t$.
After each layer, a non-linear elementwise operator $\rm ReLu(\cdot)$ is followed, i.e.
\begin{equation}
    {\rm ReLu}( h^{\phi,l}_{t,i})=\max (h^{\phi,l}_{t,i},0).
\end{equation}
The output layer will yield $\hat{\boldsymbol \mu}_t$ and $\hat{\Sigma}_t$.

\subsection{Non-affine Transform}
In this section, we describe the conditioner and transformer of the adopted transform.
\subsubsection{Conditioner}
The function $c_k(\cdot)$ is set as an additive model and implemented by an NN.
Concretely, it contains two parts: the function of $\boldsymbol z^{(k)}_{t,1:i-1}$ and the function of $\boldsymbol x_t$.
Then, $c_k(\cdot)$ is described as 
\begin{equation}
    c_k(\boldsymbol z^{(k)}_{t,1:i-1},\boldsymbol x_t;\hat{\theta}_{c_k})=c_{k,1}(\boldsymbol z^{(k)}_{t,1:i-1})+c_{k,2}(\boldsymbol x_t),
\end{equation}
where $c_{k,1}(\cdot)$ and $c_{k,2}(\cdot)$ are the two component functions.
$c_{k,2}(\cdot)$ is implemented by an NN, similar to that of $\phi(\cdot)$.
Specially, as the length of $\boldsymbol z^{(k)}_{t,1:i-1}$ changes for each dimension, it is implemented via a model named as MADE \cite{germain2015made}.

\subsubsection{Transformer}
The main idea of a spline-based NF is to implement the transform as a monotonic spline \cite{durkan2019neural}.
Each $\tau_k$ is represented as a piece-wise function which contains $M$ segments specified by $M+1$ coordinates~(knots).
The knots are obtained from the conditioner $c_k(\cdot)$ and denoted as $\{(\alpha_{k,m},\beta_{k,m})|m=0,\cdots,M \}$.
Accordingly, the transformer $\tau_k(\cdot)$ is split into $M$ segments, each of which is a simple monotonic function.
Every two nearby segments will meet at internal knots $\{(\alpha_{k,m},\beta_{k,m})|m=1,\cdots,M-1 \}$.
Specifically, we use monotonic rational-quadratic splines, which are defined by derivatives at internal knots besides the knots.
They are also derived from the conditioner $c_k(\cdot)$ and denoted as $\{\delta_{k,m}|m=1,\cdots,M-1 \}$.
We define 
\begin{align*}
    s_{k,m} = \frac{\beta_{k,m}-\beta_{k,m-1}}{\alpha_{k,m}-\alpha_{k,m-1}},
\end{align*}
\begin{align*}
    \xi(z_{t,i}^{(k-1)})=\frac{z_{t,i}^{(k-1)}-\alpha_{k,m-1}}{\alpha_{k,m}-\alpha_{k,m-1}}.
\end{align*}
The rational-quadratic function in the $m$-th bin is expressed as
\begin{align*}
\begin{split}
    r_{k,m}(\xi)=
    &\beta_{k,m-1}+\\
    &\frac{(\beta_{k,m}-\beta_{k,m-1})[s_{k,m}\xi^2+\delta_{k,m-1}\xi(1-\xi)]}{s_{k,m}+[\delta_{k,m}+\delta_{k,m-1}-2s_{k,m}]\xi(1-\xi)},
\end{split}
\end{align*}
where $\xi$ represents $\xi(z_{t,i}^{(k-1)})$.
That is, 
\begin{equation}
    \tau_k(z_{t,i}^{(k-1)})=r_{k,m}(\xi),\ {\rm if}\ z_{t,i}^{(k-1)}\in [\alpha_{k,m-1},\alpha_{k,m}].
\end{equation}
Specifically, when $z_{t,i}^{(k-1)}<\alpha_{k,0}$ or $z_{t,i}^{(k-1)}>\alpha_{k,M}$, we set $\tau_k(\cdot)$ as equivalent transform, i.e.,
\begin{equation}
    \tau_k(z_{t,i}^{(k-1)})=z_{t,i}^{(k-1)},\ {\rm if}\ z_{t,i}^{(k-1)}\in (-\infty,\alpha_{k,0}]\cup [\alpha_{k,M},\infty).
\end{equation}

As $\tau_k(\cdot)$ is monotonic, the inverse path can be computed analytically by solving a quadratic equation, i.e.,
\begin{equation}
    \xi(z_{t,i}^{(k-1)})=\frac{2C}{-B-\sqrt{B^2-4AC}},
\end{equation}
where
\begin{align*}
\begin{split}
    A = &(\beta_{k,m}-\beta_{k,m-1})(s_{k,m}-\delta_{k,m-1})\\
    &+(z_{t,i}^{(k)}-\beta_{k,m-1})(\delta_{k,m}+\delta_{k,m-1}-2s_{k,m}),
\end{split}
\end{align*}
\begin{align*}
\begin{split}
    B = &(\beta_{k,m}-\beta_{k,m-1})\delta_{k,m-1}\\
    &-(z_{t,i}^{(k)}-\beta_{k,m-1})(\delta_{k,m}+\delta_{k,m-1}-2s_{k,m}),
\end{split}
\end{align*}
\begin{align*}
    C=-s_{k,m}(z_{t,i}^{(k)}-\beta_{k,m-1}).
\end{align*}
It implicitly defines the inverse function $\tau_k^{-1}(\cdot)$. 

\section{Case study}

In this paper, we validate the proposed approach in both univariate cases~(Case 1, Case 2) and multivariate cases~(Case 3, Case 4), , which cover typical applications in probabilistic wind power forecasting. Case 1 and case 2 differ in forecast horizons. Specifically, Case 1 aims at day-ahead forecasting, whereas Case 2 focuses on forecasting within minutes to hours. Case 3 aims at characterizing the joint distribution of wind power values for various lead times, jointly. Case 4 deals with the joint distribution of wind power values at several geographical locations.
Their settings are described as follows and summarized in Table \ref{table_case}.
\begin{enumerate}
    \item Case 1: It is a day-ahead PWPF case based on GEFCom 2014 data\footnote{Available at http://blog.drhongtao.com/2017/03/gefcom2014-load-forecasting-data.html}, where numerical weather predictions~(NWPs) are taken as inputs and the predictive PDF of wind power at each time step is issued as forecast.
    \item Case 2: It is a very-short-term PWPF case where previous values of wind power generation are taken as inputs, and the predictive PDF of wind power at future time is issued as forecast.
    The horizon is set as 1 here for validation based on NREL\footnote{Available at https://www.nrel.gov/grid/wind-toolkit.html} and France wind farm data\footnote{Available at https://opendata-renewables.engie.com/explore/index}.
    \item Case 3: It is a scenario generation case based on France wind farm data, which considers temporal interdependence. 
    Specifically, we generate scenarios of future 6 time steps, which can be used in electricity market.
    \item Case 4: It is a scenario generation case based on NREL data, which considers spatial interdependence of multiple sites.
    The horizon is set as 1.
    Specifically, we choose data from 5 nearby wind farms for validation.
\end{enumerate}
As feature selection is not the focus of this paper, in Case~2, Case~3, and Case~4, the length of input features is determined by a preliminary test, which is varied from 4 to 24 and empirically set as 6.
Certainly, models may be further improved by finely selecting the features.
But it is fair for all models as they use the same input features.

\subsection{Dataset Description}

Three open datasets are used for validation, i.e., data from GEFCom~2014, NREL, and France wind farm. The GEFCom~2014 dataset provides NWPs that contain wind speeds and directions at 10-m and 100-m, and corresponding normalized wind power generation values. It is an hourly data set collected in 2012 and 2013, and contains a total of 16,800 samples. We randomly select data from 5 wind farms for experiments. The France wind farm data and NREL data are time series. 
Data from the France wind farm are collected from four wind turbines, whereas NREL data are generated by simulation at various sites. The resolution of the France wind farm data is 10-min, whereas that of the NREL data is 15-min. Specifically, we select France wind farm data collected in 2013 which contain 52355 samples, and NREL data collected in 2012 which contain 35040 samples for validation. In each case, we split $ 70\%$ of the data as a training set, $ 10\%$ as a validation set, and $ 20\%$ as a test set according to \cite{hastie2009elements}.

\subsection{Assessment Metrics}
In this paper, reliability diagrams and PI width are used to assess the reliability and sharpness of univariate predictive densities.
The comprehensive quality of predictive probability density in univariate cases is assessed by continuous ranked probability score~(CRPS) as suggested by \cite{messner2020evaluation}.
And, the quality of predictive probability density in multivariate cases is assessed by scenarios in terms of energy score~(ES) and variogram score~(VS) as suggested by \cite{pinson2012evaluating,scheuerer2015variogram}, which are allowed to measure the dependence within scenarios.
All of them are averaged over the whole test data.

\subsubsection{CRPS}
Let $F_t(y)$ denote the CDF of $Y_t$ and $ y_t $ denote the observation at time $t$. The CRPS is defined as:
\begin{equation}
    {\rm CRPS}(F_t,y_t)=\int_y(F_t(y)-\mathbbm{1}(y-y_t))^2d y,
\end{equation}
where $\mathbbm{1}(\cdot)$ is unit step function, which represents the empirical CDF of observation. 

\subsubsection{ES}
Given a set of scenarios $\{ \Tilde{\boldsymbol y}_t^{(i)} | i=1,\cdots,S \}$ and observations $\boldsymbol y_t$, the ES is defined as
\begin{equation}
    {\rm ES}=\frac{1}{S}\sum_{i=1}^S \Vert \boldsymbol y_t-\Tilde{\boldsymbol y}_t^{(i)} \Vert_2
    -\frac{1}{2S^2}\sum_{i=1}^S\sum_{j=1}^S\Vert \Tilde{\boldsymbol y}_t^{(i)}-\Tilde{\boldsymbol y}_t^{(j)} \Vert_2,
\end{equation}
where $\Vert \cdot \Vert_2$ is the $d$-dimensional Euclidean norm.
\subsubsection{VS}
Let $y_{t,i}$ and $\Tilde{y}_{t,i}$ respectively denote the $i$-th dimension of the observation $\boldsymbol y_t$ and a scenario $ \Tilde{\boldsymbol y}_t$.
The VS is defined as
\begin{equation}
    {\rm VS}=\sum_{i,j=1}^d(|y_{t,i}-y_{t,j}|^p-\mathbb{E}(|\Tilde{y}_{t,i}-\Tilde{y}_{t,j}|^p))^2,
\end{equation}
where 
\begin{align*}
    \mathbb{E}(|\Tilde{y}_{t,i}-\Tilde{y}_{t,j}|^p)\approx\frac{1}{S}\sum_{s=1}^S|\Tilde{y}_{t,i}^{(s)}-\Tilde{y}_{t,j}^{(s)}|^p.
\end{align*}
Here we set $p$ as 0.5 as suggested by \cite{scheuerer2015variogram}.

\subsection{Benchmarks}
\subsubsection{Univariate Cases}
We set both parametric and non-parametric models as benchmarks. 
For the parametric approach, we choose NN models that rely on Gaussian and logit-normal distributional assumptions, and refer to them as NN-G and NN-L respectively. They share the same basic NN structure with the proposed model.
An MDN is established, as it is allowed to model more complex distributions compared to Gaussian distributions.
As for the non-parametric approach, we include two popular distribution-free models, namely KDE \cite{zhang2016k} and quantile regression gradient boosting machine (QRGBM) \cite{LANDRY20161061} as benchmarks, since they are proved effective in the GEFCom~2014. The QRGBM is an ensemble model that iteratively fits new tree model to minimize the quantile loss.
Concretely, in the KDE, we determine the nearest 100 neighbors of each test sample and use their corresponding wind power values to estimate the predictive PDF.
In addition, the climatology model is adopted as a naive benchmark model, which estimates the predictive probability density using all training data.

\subsubsection{Multivariate Cases}

For multivariate cases, we mainly use NN-G, and NN-L as benchmark models, since they are the most popular ones. Besides, the multivariate probabilistic ensemble~(MuPEn) \cite{van2021benchmark} is adopted as a naive benchmark. It is a generalized model of the complete-history persistence, which conducts random sampling without replacement from historical scenarios for each test sample.

\subsection{Implementation Details}
\subsubsection{Univariate Cases}
The base distributions of NN-G, NN-L, and the proposed model are set as Gaussian distributions. The NN that determines shape parameters of the Gaussian distributions contains 2 hidden fully connected layers (each has 512 units). For fairness, we use the same amount of transforms (concretely, 5 transforms here) for NN-G, NN-L, and the proposed model.
All the transforms are implemented by NNs with 2 hidden fully connected layers, each of which contains 256 units. 
Such transforms in the proposed model are specified as neural spline transforms\footnote{Code is available at https://github.com/honglinwen/Conditional-normalizing-flow-for-wind-power-forecasting}, whereas they are are designed as affine transforms in the NN-G and NN-L. particularly, for NN-L, we use a sigmoid transform behind the 5 affine transforms.
All hyper-parameters are tuned by cross validation. The results on condition of different hyper-parameters are reported in the appendix.
The MDN for use contains 10 Gaussian components, both the weights and shape parameters of which are estimated by an NN.

\subsubsection{Multivariate Cases}

For multivariate cases, NN-G, NN-L, and the proposed model use the same NN architecture in univariate cases. The only difference is that we adopt the auto-regressive structure here to model the joint probability density. It is implemented based on a masked auto-encoder \cite{germain2015made} that forces each variable to only rely on the previous variables in a given order via masks.
Besides, we permute variable orders after each transform, as PDF is permutation-invariant.

%\cite{Ke2017LightGBMAH} \cite{sklearn_api}
NN-G, NN-L, and the proposed model are established via Pytorch and trained by the Adam optimizer \cite{Kingma2014AdamAM}. The learning rate is determined through a grid search and ultimately set as 1e-4. It decays 1/3 per 300 iterations. The QRGBM is implemented based on lightGBM\footnote{https://lightgbm.readthedocs.io/en/latest/}, the hyperparameters of which are set according to the winner of GEFCom~2014 \cite{LANDRY20161061}. KDE is implemented by using scikit-learn\footnote{https://scikit-learn.org/stable/}.

\section{Results and Discussion}

\begin{table}[!t]
\renewcommand{\arraystretch}{1.25}
\caption{CRPS values in Case 1 (percentage of nominal capacity).}
\label{table_gefcom}
\centering
\begin{tabular}{m{1.5cm}|m{0.95cm}<{\centering}|m{0.95cm}<{\centering}|m{0.95cm}<{\centering}|m{0.95cm}<{\centering}|m{0.95cm}<{\centering}}
\hline
 & 1 & 3 & 5 & 7  & 9 \cr
 \hline
Climatology & 19.30& 18.38& 21.36&  18.10& 18.79 \cr
\hline
NN-G & 9.45& 8.98& 8.51&  7.43& 8.48 \cr
\hline
NN-L & 9.33& 8.62& 8.88&  7.40& 8.88 \cr
\hline
QRGBM & 9.72& 8.57& 8.32&  7.62& 8.27 \cr
\hline
KDE & 10.07& 8.76& 8.64&  7.76& 8.56 \cr
\hline
MDN & 9.57& 8.58& 8.19&  7.52& 9.22 \cr
\hline
Proposed model & 9.08& 8.35& 8.14&  7.09& 8.28 \cr
\hline
\end{tabular}
\end{table}
\subsection{Case~1}

\begin{figure}[!t]
\centering
\includegraphics[width=3.4in,height=2.5in]{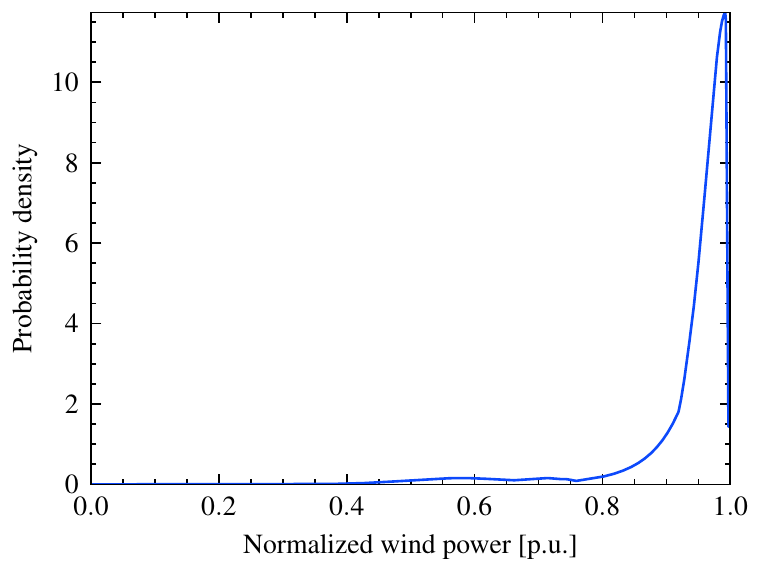}
\caption{Illustration of probability density of 100-th sample in test set.}
\label{density100}
\vspace{-1em}
\end{figure}

\begin{table*}[!t]
\renewcommand{\arraystretch}{1.25}
\caption{CRPS values in Case 2 (percentage of nominal capacity).}
\label{table_frence&nrel}
\centering
\begin{tabular}{m{1.25cm}|m{1.4cm}<{\centering}|m{1cm}<{\centering}|m{1cm}<{\centering}|m{1cm}<{\centering}|m{1cm}<{\centering}|m{1cm}<{\centering}|m{1cm}<{\centering}}
\hline
 & Climatology & NN-G & NN-L & QRGBM  & KDE & MDN &  Proposed model\cr
 \hline
France wind farm & 13.40& 1.85& 1.82&  1.92& 3.17 & 2.06 & 1.83 \cr
\hline
NREL & 21.96& 0.25& 0.25&  0.34& 2.58 & 0.46& 0.28\cr
\hline
\end{tabular}
\end{table*}

\subsubsection{CRPS}
CRPS values are presented in Table~\ref{table_gefcom}. It is seen that all the benchmark models and the proposed model outperform climatology model.
Amongst the benchmark models, KDE has slightly worse performance than others, which suggests that it is overly simplified to approximate the conditional PDF by the density of neighborhood population.
Concretely, the distribution of samples is not homogeneous, which means that more samples could be taken to better estimate the conditional PDF if the neighborhood distribution is dense.
However, once the criterion to select neighborhood samples is fixed, e.g. value of $k$ in $k$-nearest neighbors here, it cannot adaptively adjust the population, on which the conditional PDF estimation is based.
On the contrary, NN-G, NN-L, QRGBM, MDN, and the proposed model can adaptively estimate the conditional PDF/quantile by excavating the similarity of input features via parameterization or entropy measure.
It also suggests that the performance of KDE can be further improved by carefully designing such population selection criterion and making use of the similarity of neighborhood samples.
The QRGBM outperforms the NN-G and NN-L in 3/5 of cases as it is distribution-free. However, the other two cases suggest that the independent fitting in QRGBM may accumulate errors.
MDN is comparable with NN-G and NN-L, which may be explained as that it is harder to estimate weights and component distributions jointly in the MDN compared to NN-G and NN-L (where only shape parameters are required to be estimated).
The weight of MDN, namely $\pi_i(\boldsymbol x_t)$ can be interpreted as the possibility that samples will fall in the $i$-th mixing distribution. Then, by increasing the number of distributions to infinite, the approximation capability will accordingly increase, i.e.,
\begin{align*}
    f(\boldsymbol y_t|\boldsymbol x_t)=\int \pi(\boldsymbol x_t)f(\boldsymbol y_t;\boldsymbol \mu_i(\boldsymbol x_t),\boldsymbol \Sigma_i(\boldsymbol x_t))d \pi(\boldsymbol x_t).
\end{align*} However, MDNs often occur mode collapse and training instability when the number of mixing components is large or the dimension of variables is high. In this case, we investigate the number of mixing components via a grid search, concretely, 3, 10, 20, 50, 100. It turns out that the training of MDN is unstable even for 20 mixing components.
Obviously, the proposed model exceeds benchmarks in all cases.

The comparison between NN-L and NN-G shows that the logit-normal transform may deteriorate performance at times.
It reveals that the logit-normal distributional assumption may not hold sometimes, although the realizations of random variable are forced to fall into the physically defined interval.
We present the predictive probability density of the proposed model at a selected time in Fig.~\ref{density100}.
As illustrated, the PDF derived by the proposed model are more flexible than specific families of distributions because the proposed model is free of any distributional assumptions.
In addition, the proposed model has 1.7 million trainable parameters, which are comparable to those of NN-G and NN-L, i.e., 1.6 million trainable parameters.
This means that the proposed model can flexibly model different wind power distribution characteristics on condition of predicted wind speeds, with increased but affordable complexity.
 
\begin{figure}[!t]
\centering
\includegraphics[width=3.4in,height=2.5in]{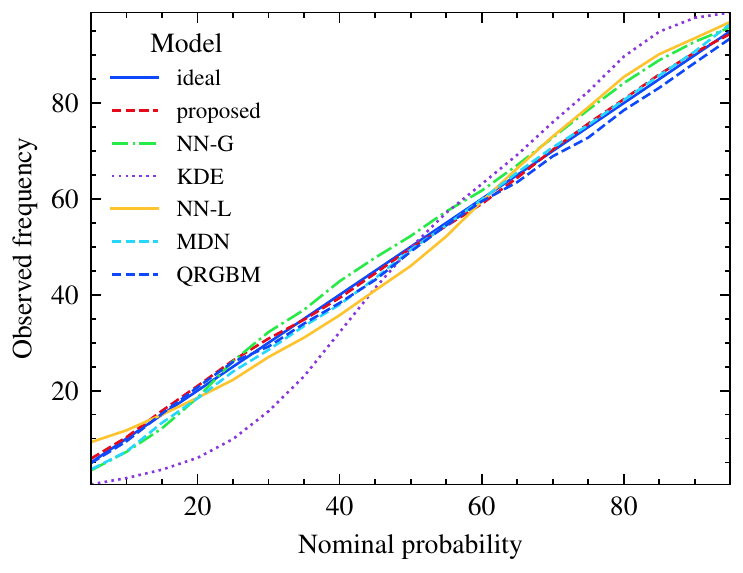}
\caption{Reliability diagram of forecasts at wind farm 1.}
\label{reliability}
\end{figure} 

\begin{figure}[!t]
\centering
\includegraphics[width=3.4in,height=2.5in]{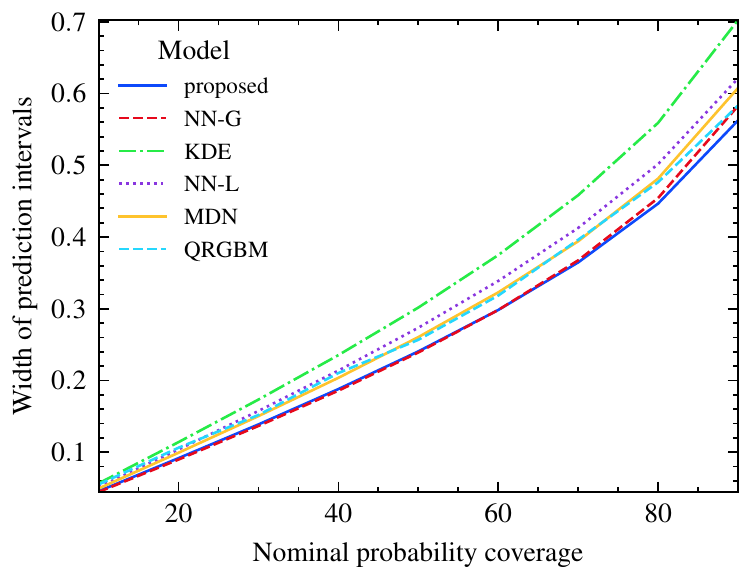}
\caption{Width of PI at wind farm 1.}
\label{width}
\end{figure} 

\begin{figure*}[!t]
\centering
\includegraphics[width=7.25in]{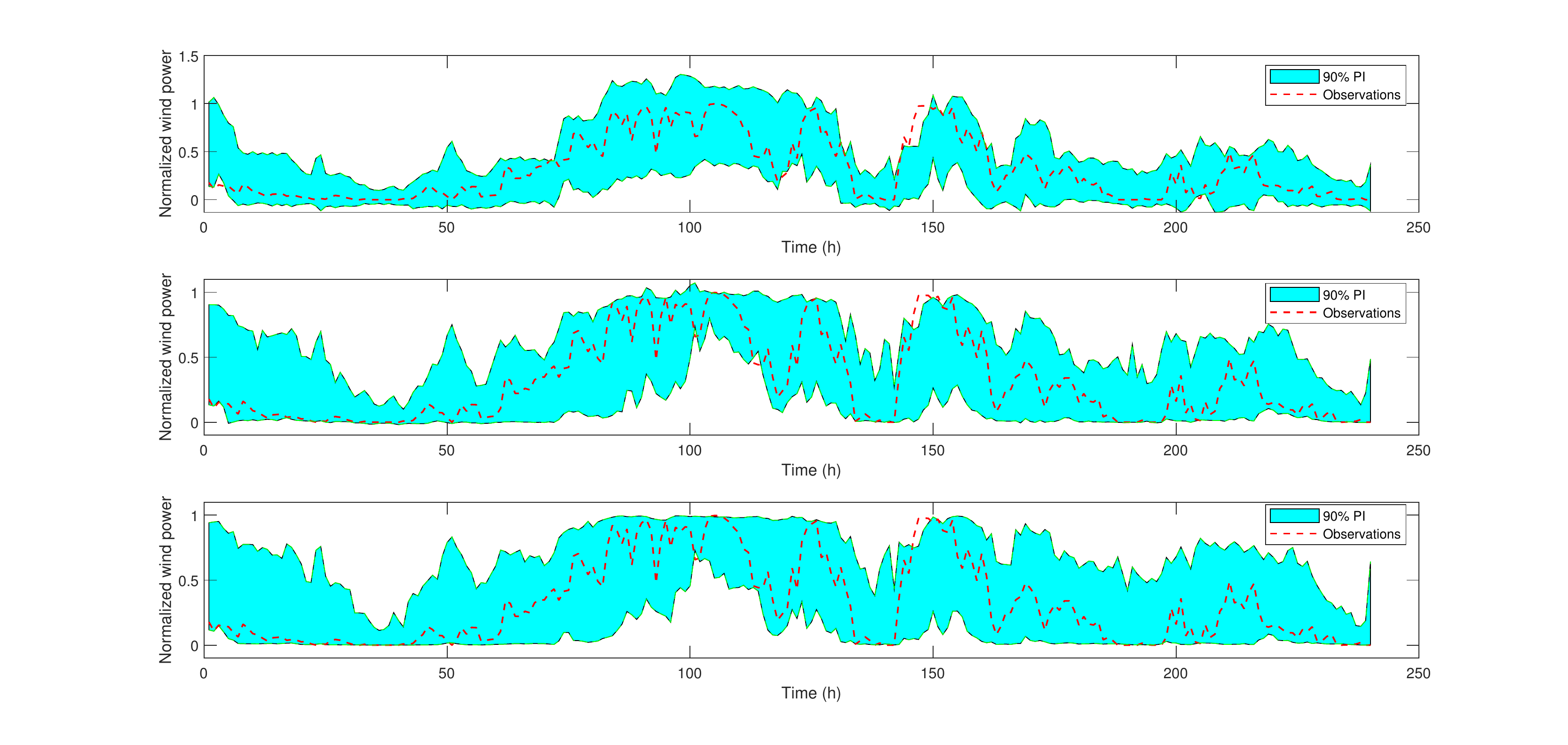}
\caption{$90\%$ PI of the proposed model of 10 days at wind farm 1, top: NN-G, middle: proposed, bottom: NN-L.}
\label{pi}
\end{figure*}

\begin{table}[!t]
\renewcommand{\arraystretch}{1.25}
\caption{ES values in Case 3 and Case 4 (percentage of nominal capacity).}
\label{table_es}
\centering
\begin{tabular}{m{1.25cm}|m{1cm}<{\centering}|m{1cm}<{\centering}|m{1cm}<{\centering}|m{1cm}<{\centering}}
\hline
 & MuPEn & NN-G & NN-L &  Proposed model\cr
 \hline
Case 3 & 33.73& 9.32& 9.20& 9.20 \cr
\hline
Case 4 & 51.95& 2.68& 2.64& 2.44\cr
\hline
\end{tabular}
\end{table}

\begin{table}[!t]
\renewcommand{\arraystretch}{1.25}
\caption{VS values in Case 3 and Case 4.}
\label{table_vs}
\centering
\begin{tabular}{m{1.25cm}|m{1cm}<{\centering}|m{1cm}<{\centering}|m{1cm}<{\centering}|m{1cm}<{\centering}}
\hline
 & MuPEn & NN-G & NN-L &  Proposed model\cr
 \hline
Case 3 & 0.3842& 0.2711& 0.2377& 0.2424 \cr
\hline
Case 4 & 0.6303& 0.0634& 0.0524& 0.0446\cr
\hline
\end{tabular}
\end{table}

\subsubsection{Reliability and Sharpness}
The reliability diagram and PI width for wind farm 1 are illustrated in Fig.~\ref{reliability} and Fig.~\ref{width}.
It turns out that QRGBM and the proposed model achieve the best performance in reliability, which are close to the ideal case.
Strictly speaking, it is unfair to compare a bunch of independently trained QR models with a single model that derives the whole distribution, as the computational cost of QR for a single quantile is much larger.
Nevertheless, the proposed model still achieves comparable reliability, which confirms its performance.
By contrast, the reliability diagrams of NN-G, NN-L, MDN, and KDE deviate from the ideal to a certain degree.
The deviation of NN-G and NN-L cannot be totally mitigated, since the families of distribution they define mismatch the real underlying distribution.
Results suggest that the superiority of the proposed model goes beyond the distribution-free property compared to the QR and KDE-based methods, by offering an efficient and continuous conditional modeling approach.

Fig.~\ref{width} demonstrates that the proposed model provides the shortest PI at all nominal levels.
However, the performance of NN-G in width of PI is comparable to that of the proposed model, whereas the PI width of NN-L is much wider.
For illustration, we present $90\%$ PI of the NN-G, proposed model, and NN-L of 10 days in the top, middle, and bottom subplots of Fig.~\ref{pi}.
As shown, the PIs of NN-G violate the bounds of wind power to a large extent, revealing probability leakage issue, while PIs of the proposed model and NN-L are more realistic.
Besides, it is demonstrated that PIs of NN-L are sometimes unnecessarily wide.
For example, between 200-h and 250-h, the upper bound of NN-L is larger than that of NN-G and the proposed model.
Indeed, both the NN-L and the proposed model can be considered as models derived from the NN-G by applying transforms.
Indeed, the logit-normal transform in the NN-L applies to all NWP conditions indifferently, whereas the spline transform of the proposed model is specified by NWPs.
This explains the sacrifice of NN-L in PI width, which is a side-effect when forcing the realizations within the boundaries.

\begin{table*}[!t]
\renewcommand{\arraystretch}{1.25}
\caption{The training time of models in Case~1 (seconds).}
\label{time_models}
\centering
\begin{tabular}{m{1.25cm}|m{0.95cm}<{\centering}|m{0.95cm}<{\centering}|m{0.95cm}<{\centering}|m{0.95cm}<{\centering}|m{0.95cm}<{\centering}|m{0.95cm}<{\centering}}
\hline
Models & NN-G & NN-L & QRGBM & KDE & MDN  & Proposed \cr
 \hline
Time & 56& 56& 44&  14& 63 & 78\cr 
\hline
\end{tabular}
\end{table*}

\subsection{Case~2}

We present the CRPS values of Case~2 in Table~\ref{table_frence&nrel}.
As with Case 1, all models are superior to the climatology model.
The performance of NN-G, NN-L, QRGBM, and the proposed model are demonstrated to be comparable.
The gap of performance between the KDE and others is enlarged compared to Case~1, because of higher dimension of input features which raises issues for $k$-nearest neighbors.
Comparing the results of KDE, QRGBM, MDN, and the proposed model with the results of NN-G and NN-L, we can infer that the Gaussian and logit-normal distributional assumptions are fairly adequate in very-short term PWPF.
This may be due to the fact that the structure of temporal interdependence over a short period of time is simpler than the interdependence spanning several hours.

\subsection{Case~3 and Case~4}
The ES and VS values are presented in Table~\ref{table_es} and Table~\ref{table_vs}.
All of NN-G, NN-L, and the proposed model outperform MuPEn, since the MuPEn draws samples from the empirical unconditional distribution whereas other models draws samples from the estimated conditional distributions.
Except for the MuPEn, the ES and VS values in Case~3 are larger than those in Case~4, which indicates larger uncertainty in Case~3.
This is is caused by the increase in generation uncertainty as forecasting horizon increases.
In both cases, NN-L and the proposed model exceed NN-G, which suggests the limited capability of the Gaussian distributional assumption in complex and high dimensional cases.
Besides, the performance of NN-L and the proposed model is comparable in Case~3, but differs in Case~4, which suggests that spatial interdependence is more complex.

\subsection{Discussion on the Base Distribution}
Theoretically, the base distribution modeled by $\mathcal{G}$ can be set as any distribution.
By learning the transforms, one can still obtain the estimation of desired distribution.
But it means that one needs to estimate the transforms in a relatively large space, if the base distribution is considerably different from the underlying distribution.
For illustration, we consider the wind farm 1 in Case 1 and set the base distribution as a standard Normal distribution $\mathcal{N}(0,1)$.
In theory, this will not make a lot differences in estimated distributions, since the standard Normal distribution can be transformed to any Gaussian distribution by using an affine transform.
But compared to the proposed model, this setting implies a more complex task, i.e., the flow model requires to estimate such affine transform besides the non-affine transform.
In the experiments, the CRPS value under the condition of standard Normal base distribution turns out as 14.9, which is much larger than that of the proposed model, i.e., 9.22.
In other words, the estimation of base distribution if of significance in the view of practice.

\subsection{Discussion on Transforms}
In this paper, we set the transformers as rational-quadratic splines. In fact, other splines could also be considered as long as the transforms are invertible, for instance linear and cubic splines \cite{muller2019neural,durkan2019cubic}. However, it is hard to compute the inverse path of high-degree spline based transforms. As suggested by \cite{durkan2019neural}, calculating the inverse path of a cubic spline is prone to numerical instability. On the other hand, it is required that transforms are flexible enough, which translates into saying that there is a trade-off between complexity and flexibility. The adopted rational-quadratic spline based flows are more flexible than linear and quadratic spline based flows.
We use linear and cubic spline based transforms for a comparative study in Case 1 and Case 3. The CRPS values for linear and cubic spline based CNF models on wind farm 1 in Case 1 are respectively 9.34 and 9.11. The ES values for linear and cubic spline based CNF models in Case 3 are respectively 9.28 and 9.22. That is, the performance of the rational-quadratic based flow is superior to that of the linear spline based flow and comparable to the cubic spline based flow. Certainly, more advanced normalizing flow models could be used.

\subsection{Distribution-free vs Distributional Assumption}
In the case study, QRGBM, KDE, and the proposed model are distribution-free, whereas NN-G and NN-L rely on distributional assumptions.
Compared to NN-G and NN-L, the proposed model has increased but affordable complexity due to its spline operation.
Meanwhile, the increased complexity enables the proposed model to obtain different wind power distributions on condition of different predicted wind speeds.
Compared to QRGBM and KDE, the proposed model is superior in efficiently modeling whole conditional PDFs.
In addition, case studies show that distribution-free methods are not overwhelmingly superior to models with distributional assumptions.
Concretely, NN-G and NN-L rival QRGBM and KDE in several cases.
And in Case~2, the performance of NN-L is comparable to that of the proposed model, which means these distributional assumptions are adequate in very-short-term PWPF.
But when it comes to applications with more uncertainty and more complex interdependence, the proposed approach always achieves a satisfactory performance with an acceptable computational cost.
Indeed, it has been reported in \cite{dumas2021deep} that the distribution-free integration-based NF model is comparable to an affine NF model that is equivalent to NN-G.
We infer that it is resulted from the difficulty in training the integration-based NF.
Intuitively, it will take more effort to find the desired transform in a larger function space.
This also reveals the trade-off between complexity and flexibility in modeling distributions.

\subsection{Training Time}

The training time of all models in Case~1 is presented in Table~\ref{time_models}, we report the training time of NN-based models in 1000 iterations and 199 independent quantiles of QRGBM.
It shows that the training time of the proposed model is comparable to that of commonly used NN-G, which is affordable.
In general, the training time of the proposed model is governed by the number of transforms and the number of hidden units.
With more transforms and hidden units, the training time will increase.
However, it still costs time to generate scenarios for high-dimensional multivariate forecasting.

\section{Conclusions}
The approach for probabilistic wind power forecasting described in this paper, based on conditional spline normalizing flow, offers a number of advantages with respect to the existing. It directly estimates the conditional probability density and does not require any assumption on the distributions involved. In addition, it is applicable to both univariate and multivariate PWPF, with high efficiency in terms of both modeling and computing. Our case-study applications based on open datasets confirmed the interest of the approach and its wide applicability for wind power applications.

Parameters are assumed fixed in this paper; therefore it is still required to explore how to estimate the parameters in an online learning fashion. Besides, the time for scenario generation is costly when dimension increases, so we will focus on finding more efficient methods in the future.

%Appendix one text goes here.
\appendix
%Please refer to the appendix document.

\begin{table}[!t]
\renewcommand{\arraystretch}{1.25}
\caption{CRPS under different steps of transforms (percentage of nominal capacity).}
\label{table_transform}
\centering
\begin{tabular}{m{1.35cm}|m{0.95cm}<{\centering}|m{0.95cm}<{\centering}|m{0.95cm}<{\centering}|m{0.95cm}<{\centering}|m{0.95cm}<{\centering}}
\hline
Number of Transforms & 1 & 2 & 3 & 4  & 5 \cr
 \hline
CRPS & 9.93& 9.51& 9.23&  9.25& 9.22\cr 
\hline
\end{tabular}
\end{table}

\begin{table}[!t]
\renewcommand{\arraystretch}{1.25}
\caption{CRPS under different sizes of hidden units (percentage of nominal capacity).}
\label{table_units}
\centering
\begin{tabular}{m{1.25cm}|m{0.95cm}<{\centering}|m{0.95cm}<{\centering}|m{0.95cm}<{\centering}}
\hline
Number of Units & 64 & 256 & 512  \cr
 \hline
CRPS & 9.22& 9.08 & 9.37\cr 
\hline
\end{tabular}
\end{table}

\begin{table}[!t]
\renewcommand{\arraystretch}{1.25}
\caption{CRPS under different number of knots (percentage of nominal capacity).}
\label{table_knots}
\centering
\begin{tabular}{m{1.25cm}|m{0.95cm}<{\centering}|m{0.95cm}<{\centering}|m{0.95cm}<{\centering}|m{0.95cm}<{\centering}}
\hline
Knots & 5 & 10 & 20 & 50\cr
 \hline
CRPS & 9.25& 9.08& 9.19&  9.20\cr 
\hline
\end{tabular}
\end{table}
% you can choose not to have a title for an appendix
% if you want by leaving the argument blank
%\section{}
%Appendix two text goes here.

\subsection{Selection on Hyperparameters}
To empirically determine the hyperparameters, we conduct a preliminary test to validate the influence of number of transforms, number of units, and number of knots by studying variants of Case~1.
Specifically, we take wind farm 1 as an example, and present results of several case settings. 
\subsubsection{Number of Transforms}
In this case, we set the number of hidden units in transform as 64, the number of knots as 10, and vary the number of transforms from 1 to 5.
The corresponding results are shown in Table~\ref{table_transform}.
It can be seen that the CRPS is relatively larger when we use only few transforms.
Consequently, the model is small, which results in limited capability of fitting ultimate transform and shape parameter function of base distribution.
After reaching at 3 transforms, the gain of increasing transforms is relatively low, which suggest the capability is enough.
Besides, increasing transforms means increasing layers of deep neural network, whose training procedure might become difficult when the model is considerably deep.
\subsubsection{Number of Hidden Units}
Here we fix the number of transforms as 5, the number of knots as 10, and adjust the number of hidden units as 64, 256, and 512.
Results are presented in Table~\ref{table_units}.
It shows that the fitting capability of NN in each transform is influenced by the number of hidden units.
The capability is limited when the number of hidden units is few.
But it might overfit the data if the number of hidden units is considerable.
\subsubsection{Number of Knots}
In this case, we fix the number of layers as 5 and the number of hidden units as 256, and look into the influence of knots by varying the number.
We set it as 5, 10, 20, and 50 respectively, whose results are shown in Table~\ref{table_knots}.
As we increase the number of knots, the CRPS first decreases and then increases.

% use section* for acknowledgment

\section*{Acknowledgment}

%This work is supported by Shanghai Sailing Program (19YF1423700) and Key Project of Shanghai Science and Technology Committee (18DZ1100303).
This work was performed during a research stay at the Technical University of Denmark.
The authors would like to appreciate China Scholarship Council (NO. 202006230261) and Shanghai
Sailing Program (19YF1423700). The research leading to this work is being carried out as a part of the Smart4RES project (European Union’s Horizon 2020, No. 864337). The sole responsibility of this publication lies with the authors. The European Union is not responsible for any use that may be made of the information contained therein.

% Can use something like this to put references on a page
% by themselves when using endfloat and the captionsoff option.
\ifCLASSOPTIONcaptionsoff
  \newpage
\fi

% trigger a \newpage just before the given reference
% number - used to balance the columns on the last page
% adjust value as needed - may need to be readjusted if
% the document is modified later
%\IEEEtriggeratref{8}
% The "triggered" command can be changed if desired:
%\IEEEtriggercmd{\enlargethispage{-5in}}

% references section

% can use a bibliography generated by BibTeX as a .bbl file
% BibTeX documentation can be easily obtained at:
% http://mirror.ctan.org/biblio/bibtex/contrib/doc/
% The IEEEtran BibTeX style support page is at:
% http://www.michaelshell.org/tex/ieeetran/bibtex/
%\bibliographystyle{IEEEtran}
% argument is your BibTeX string definitions and bibliography database(s)
%\bibliography{IEEEabrv,../bib/paper}
%
% <OR> manually copy in the resultant .bbl file
% set second argument of \begin to the number of references
% (used to reserve space for the reference number labels box)
\bibliographystyle{IEEEtran}
% argument is your BibTeX string definitions and bibliography database(s)
\bibliography{IEEEabrv,mylib}

% biography section
% 
% If you have an EPS/PDF photo (graphicx package needed) extra braces are
% needed around the contents of the optional argument to biography to prevent
% the LaTeX parser from getting confused when it sees the complicated
% \includegraphics command within an optional argument. (You could create
% your own custom macro containing the \includegraphics command to make things
% simpler here.)
%\begin{IEEEbiography}[{\includegraphics[width=1in,height=1.25in,clip,keepaspectratio]{mshell}}]{Michael Shell}
% or if you just want to reserve a space for a photo:

%\begin{IEEEbiography}{Michael Shell}
%Biography text here.
%\end{IEEEbiography}

% if you will not have a photo at all:
%\begin{IEEEbiographynophoto}{John Doe}
%Biography text here.
%\end{IEEEbiographynophoto}

% insert where needed to balance the two columns on the last page with
% biographies
%\newpage

%\begin{IEEEbiographynophoto}{Jane Doe}
%Biography text here.
%\end{IEEEbiographynophoto}

% You can push biographies down or up by placing
% a \vfill before or after them. The appropriate
% use of \vfill depends on what kind of text is
% on the last page and whether or not the columns
% are being equalized.

%\vfill

% Can be used to pull up biographies so that the bottom of the last one
% is flush with the other column.
%\enlargethispage{-5in}

% that's all folks
\end{document}